\documentclass[aps, prb, unsortedaddress,floatfix, nofootinbib, twocolumn, superscriptaddress,floatfix, longbibliography]{revtex4-2}
\usepackage{graphicx, color, soul}
\usepackage{amssymb}
\usepackage[usenames,dvipsnames, pdftex]{xcolor}
\usepackage{ulem}
\usepackage{bm}
\usepackage[mathscr]{eucal}
\usepackage{physics}
\usepackage{mhchem}

\usepackage[colorlinks, breaklinks, 
linkcolor=NavyBlue,
citecolor=NavyBlue,
urlcolor=NavyBlue]{hyperref}

\usepackage[all]{hypcap}

\newcommand{\msr}{\mathscr}
\newcommand{\eq}[1]{\begin{align}#1\end{align}}

\begin{document}

\author{Ruchira V Bhat}
\affiliation{Department of Physics, Indian Institute of Technology Bombay, Mumbai 400076}

\author{Soumya Bera}
\affiliation{Department of Physics, Indian Institute of Technology Bombay, Mumbai 400076}

\title{Distinguishing dynamical quantum criticality through local fidelity distances}%

\date{\today}

\begin{abstract}
Using local quantum fidelity distances, we study the dynamical quantum phase transition in integrable and non-integrable one-dimensional Ising chains.  
Unlike the Loschmidt echo, the standard measure for distinguishing between two quantum states to describe the dynamical quantum phase transition, the local fidelity requires only a part of the system to characterize it.
The non-analyticities in the quantum distance between two subsystem density matrices identify the critical time and the corresponding critical exponent reasonably well in a finite-size system.  
Moreover, we propose a distance measure from the upper bound of the local quantum fidelity for certain quench protocols where the entanglement entropy features oscillatory growth in time.
This local distance encodes the difference between the eigenvalue distribution of the initial and quenched subsystem density matrices and quantifies the critical properties.
The alternative distance measure could be employed to examine the dynamical quantum phase transitions in a broader range of models, with implications for gaining insights into the transition from the entanglement perspective. 
\end{abstract}

\maketitle

\section{Introduction}
Conventional symmetry-breaking quantum phase transitions are described within the Landau-Ginzburg paradigm, where the free energy density becomes nonanalytic at the transition point \cite{secondorder1,secondorder3}. 
Unlike the symmetry-breaking quantum phases, the topological phases are understood from the response of the wavefunction under small adiabatic changes of the Hamiltonian and are  characterized by topological invariants \cite{Topology1,Topology2,Topology3}. 
On the contrary, an out-of-equilibrium phase transition signaled by the non-analyticities in time, such as the dynamical quantum phase transition (dQPT), is characterized by the zeros of the boundary partition function in the complex time plane \cite{Heyl_DQPT_review, Heyl_survey,DQPT_finite_temp, DQPT_finite_temp2}.
The boundary partition function is defined via the Loschmidt amplitude $\mathcal{L}(z)=\bra{\psi_{0}}e^{-z H} \ket{\psi_{0}}$, where $z=it$ is the imaginary time, and $\psi_0$ is the initial state; the nonanalyticities of the boundary partition function translates as the singularities of the rate function,
\eq{ \lambda(t) = -\lim_{L \rightarrow \infty} \frac{1}{L} \log |\bra{\psi_{0}}e^{-i H t} \ket{\psi_{0}}|^{2} \label{returnprobability},}
which defines the critical time $t^*$ of such phase transition~\cite{Heyl_DQPT_review,Heyl_DQPT,Heyl_scaling_dqpt,Heyl_dqpt_TFIM,dqpt1,dqpt2,dqpt3,DQPT_weyl}. The $\lambda(t)$ obeys dynamical scaling laws in accordance to conventional phase transition and provided the initial framework to analyze, predict, and classify the dQPTs'.

For example, the dQPT observed in one-dimensional integrable and non-integrable transverse field Ising models during
a global quench between paramagnetic and ferromagnetic phases fall in the Ising universality class, where $\lambda(t)$ follows a  power law scaling in time with an exponent, $\nu \backsimeq 1$ \cite{quasilocal,Heyl_scaling_dqpt}. 
Conversely, the dQPT in the two-dimensional Ising model belongs to a different universality class that is characterized by logarithmic scaling of $\lambda(t)$ in time~\cite{Heyl_scaling_dqpt}.
%
Moreover, in systems with underlying equilibrium topology, the dQPT is characterized not only by $\lambda(t)$ but also by a dynamical topological order parameter~\cite{DQPT_top1, DQPT_top2, DQPT_top3, DQPT_top4, DQPT_top5,DQPT_top6, DQPT_top7, DQPT_top8,DQPT_top9}. 
Even though the condition of observing dQPT was initially associated with the presence of an equilibrium critical point~\cite{DQPT_underlying_equilibrium1,DQPT_underlying_equilibrium2,Heyl_DQPT,Heyl_dqpt_TFIM}, it has now become accepted that dQPTs are fundamentally non-equilibrium phenomena that can occur without any equilibrium counterpart~\cite{DQPT_pure1,DQPT_pure2,DQPT_pure3,DQPT_pure4,DQPT_pure5}. 
%
\begin{figure*}[!tbh]
    \centering
    \includegraphics[width=0.8\textwidth]{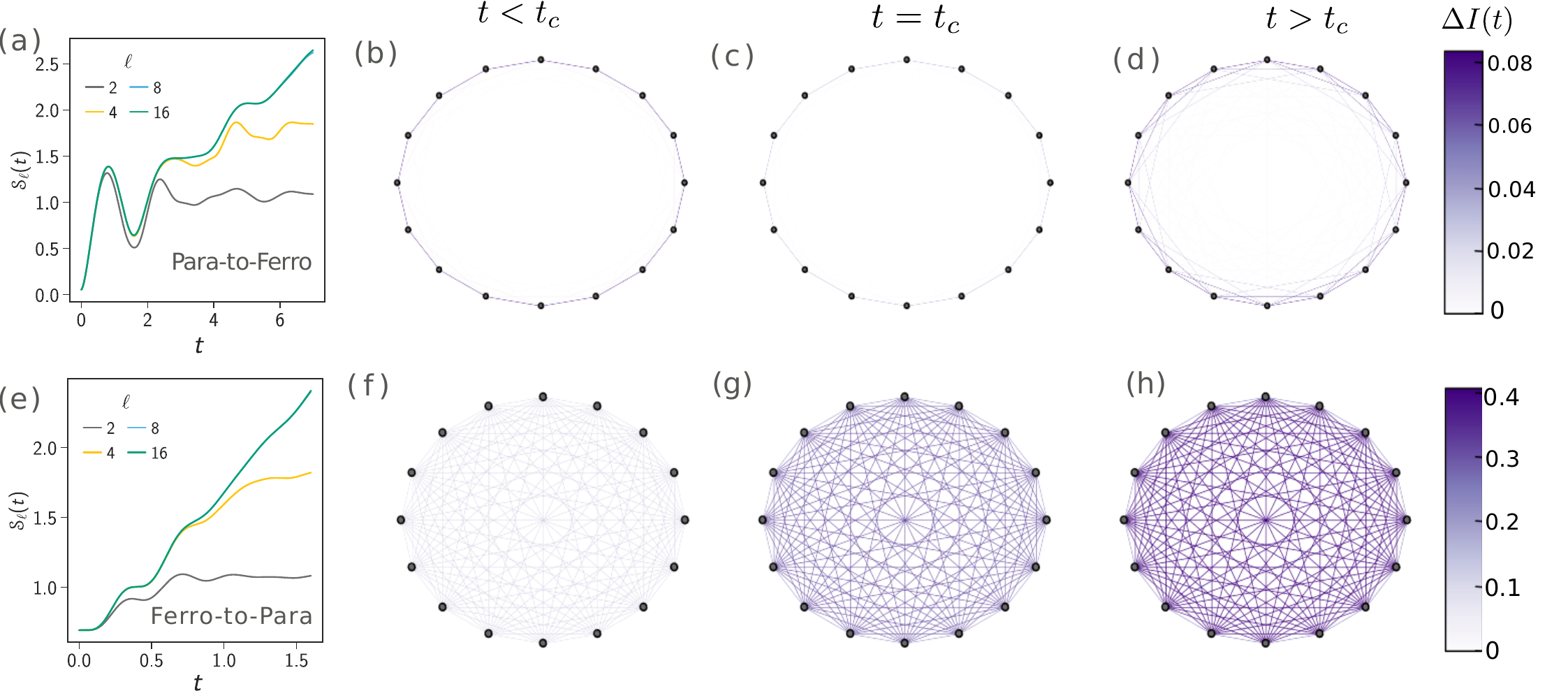}
    \caption{{\bf Entanglement entropy and mutual information for TFIM.}~The entanglement entropy with oscillatory and nearly linear growth with time under quench-I and quench-II (see text for the definition of the quench protocols) for the transverse field Ising model is shown in (a) and (e). (b), (c), and (d) represent the change in mutual information between two spins during quench-I at three instants of time, $t_{c}$ is the critical time. (f), (g) and (h) show the mutual information change for the opposite quench. The change in long-range correlations observed for quench-II is almost fives times more than that for quench-I.}
    \label{MI_Int}
\end{figure*}

However, relying solely on $\lambda(t)$ and the topological order parameter presents challenges in fully understanding dQPTs, primarily for two reasons.
Firstly, these observables fall short in distinguishing dQPTs resulting from distinct quenches that exhibit apparent differences in entanglement entropy and the dynamics of local correlations~\cite{Entanglement_view, SSH_DQPT, DQPT_entanglement2,DQPT_entanglement3,DQPT_entanglement4}. 
The entanglement echo, a rate function analog that
encodes the deviation of the entanglement ground state
from the instantaneous entanglement state, partially addresses the abovementioned issue by distinguishing
dQPTs with and without oscillations in the entanglement entropy growth in time \cite{DQPT_entanglement3}.
Secondly, global observables are usually challenging to access in generic experiments.

Recently, \textcite{DQPT_trappedions} simulated the temporal dynamics of an interacting transverse field Ising model in a finite number of trapped ions. The appearance of dQPT was revealed through the non-analyticities in the rate function within the degenerate ground state manifold, $\lambda_{p}(t) = - \mathrm{min}_{p~\in~(\uparrow,\downarrow)}\log |\bra{\psi_{p}} e^{-i H t}\ket{\psi_{p}}|^{2}/L$, where $\ket{\psi_{\uparrow(\downarrow)}}$ is the symmetry breaking doubly degenerate state and in the thermdynamic limit~$L \rightarrow \infty$, the finite size $\lambda_{p}(t) $ converges to the rate function~\eqref{returnprobability}. This study further establishes a direct connection between this microscopic probability and the macroscopic observables by demonstrating a repeated crossover in magnetization dynamics between positive and negative sectors, along with entanglement production near the critical time. 

There are also a few theoretical advances in defining local observables to characterize dQPT's. For instance, a real-space local effective free energy~\cite{Local_measures} and a quasilocal string measure~\cite{quasilocal,quasilocal2}, have been introduced as local observables to characterize dQPTs.
The real-space local effective free energy, $\lambda_{M}(t) = - \log|\mathcal{L}_{M}(t)|^{2}/M$ where $~|\mathcal{L}_{M}(t)|^{2} = \bra{\psi(t)}P^{z}_{M}\ket{\psi(t)}$, $M$ is the subsystem size in the real space and $P^{z}_{M} = \frac{1}{L} \sum_{i=1}^{L} \frac{1}{2^{M}} \Pi_{i}^{i+M} (I_{i} + \sigma^{z}_{i})$ is the projector, can be thought of as the rate function corresponding to a smaller part of the system, such that in the limit $M\rightarrow L$,~$\lambda_{M}(t)\approx\lambda(t)$. 
Furthermore, the momentum local counterpart  of $\lambda_{M}(t)$, which is calculated in terms of
two point single particle correlations in the integrable limit facilitate the experimental detection of dQPT, particularly in scenarios where measurement of local observables in $k$ space are involved~\cite{Local_measures,quasilocal}.
However, it is not evident whether these local order parameters will be sufficient to distinguish dQPTs with different behavior of entanglement entropy and subsystem correlations in time. 

Here, we propose two local quantum distance measures, quantum reduced fidelity distance (qRFD) and minimum reduced fidelity distance (mRFD), to characterize dQPTs observed in integrable and non-integrable spin models.
We investigate two quench protocols that show distinct behaviors in entanglement entropy growth: one showing oscillatory growth and the other exhibiting
linear growth near the critical time as shown in Fig.~\ref{MI_Int}(a) and (e).   
The qRFD quantifies the distinguishability between initial and quenched reduced density matrices and displays nonanalytic behavior associated with the corresponding dQPT, regardless of the quench protocol. 
In contrast, the mRFD differentiates between the eigenvalue distribution of initial and quenched reduced density matrices, characterizing the oscillatory entanglement growth, and therefore the dQPT.

The computation of mRFD requires only the eigenvalues of the reduced density matrices. 
It implies that for integrable models (with and without non-trivial underlying momentum space topology), the single particle correlation matrix, which scales with the linear dimension of the system, is sufficient to find the relevant scaling exponents. 

\section{Local distance measures}\label{qRFD and mRFD}
\subsection{Quench protocols}\label{quenchprotocols}
We consider two quench protocols that shows dQPT in spin and fermionic chains.
The first type, quench-I, from the paramagnetic (trivial) to ferromagnetic (topological) phase, shows oscillations (see Fig.~\ref{MI_Int}(a)) in the growth of entanglement entropy at initial times and concomitantly avoided level crossings in the reduced density spectrum near the transition time (see Section \ref{mrfd_and_Correlations} for details). 
The opposite one, quench-II, exhibits a linear increase in entanglement $\msr{S}(t)=-\mathbf{Tr} \left[ \rho^\ell(t)\log\rho^\ell (t)\right] $, where $\rho^\ell(t) $ is the time dependent subsystem $\ell$ density matrix,  as seen in Fig.~\ref{MI_Int}(e) with time, and a gap in the reduced density spectrum.
A further differentiation can be made between these two quenches by understanding the spatial entanglement structure after the quench, which we discuss in the following.
\subsubsection*{Local entanglement correlation} 
The spatial entanglement structure is fundamentally different depending on the direction of the quench. 
We quantify these correlations through mutual information $I(A,B) = \msr{S}(A) + \msr{S}(B) - \msr{S}(A~U~B)$, where $A$ and $B$ are two subsystems with $\msr{S}(A)$ being the entanglement entropy of subsystem ${A}$~\cite{Mutual_Information}.
Figure~\ref{MI_Int} shows the change in mutual information between two spins at sites $i,j$; $\Delta I(t)=I_{i,j}(t)-I_{i,j}(0)$, after quench-I((b), (c) and (d)) and quench-II ((f), (g) and (h)) at three instants of time.

We observe the emergence of long-range correlations in the ferro-to-para quench. This observation highlights the necessity of employing measures beyond the local subsystem distances to probe such correlations at longer times. Conversely, when the spatial entanglement remains local, we show that a minimum local distance enables the characterization of dQPTs at finite times for a finite-size system. This approach has practical value, particularly in experimental investigations.
\subsection{Quantum reduced fidelity}
The Loschmidt amplitude,~$\mathcal{L}(t) = \bra{\psi_{0}}\ket{\psi(t)}$, which outlines the rate function~\eqref{returnprobability}, is the quantum fidelity.
For a quenched system, the quantum fidelity,
$F(t) =\left(\mathbf{Tr}\sqrt{\sqrt{\rho_{0}}\rho(t)\sqrt{\rho_{0}}}\right)^{2}$, measures the distinguishability between the initial and quenched state with density matrices $\rho_{0}$  and $\rho(t)$ respectively~\cite{Fidelity_review,Fidelity_topo}. 
Similar to rate function, the quantum fidelity is also a global observable that requires access to the entire many-body wave function.

We define a local quantum reduced fidelity for a subsystem size $\ell$ as,  
\eq{F_{\ell}(\rho_{0}^{\ell}, \rho^{\ell}(t)) =  \left(\mathbf{Tr}\sqrt{\sqrt{\rho_{0}^{\ell}}\rho^{\ell}(t)\sqrt{\rho_{0}^{\ell}}}\right)^{2}. \label{quantumreducedfidelity}} 
It corresponds to the fidelity between initial $(\rho_{0}^{\ell})$ and time evolved $(\rho^{\ell}(t))$ reduced density matrices with $\rho^{\ell} = \mathbf{Tr}_{L-\ell} \rho$, $\rho$ being the total density matrix, $\ell$ is the subsystem size and  $L$, the size of the system. 
It is known that, in the context of equilibrium quantum phase transitions, the quantum reduced fidelity of a subsystem size as small as two sites characterize the critical point and exponent~\cite{red_fid1,red_fid2,red_fid3,red_fid4}.
The corresponding quantum distance is defined as,
\eq{d_{\ell}^{q}(t) = -\frac{1}{\ell} \log(F_{\ell}(\rho_{0}^{\ell}, \rho^{\ell}(t)))\label{qRFD},}
a local observable defined from the quantum reduced fidelity of the subsystem $\ell$.
Similar to quasilocal string observable \cite{quasilocal} and real-local effective free energy \cite{Local_measures}, for large $L$ and as $\ell \rightarrow L$, the local distance also approaches the rate function $d_{\ell}^{q}(t) = \lambda(t)$.
Here we show that this quantum reduced fidelity distance  exhibits non-analyticities and obeys scaling laws near the critical point of dQPTs occurring under different quenches in integrable and non-integrable models.

\subsection{Minimum reduced fidelity}
The entanglement entropy plays a crucial role in identifying and characterizing dQPTs, and is discussed in sec.~\ref{quenchprotocols}.
Additionally, the observation that the quantum reduced fidelity is bounded by the fidelity of the associated diagonal states \cite{Classical_fidelity} motivates us to develop a distance measure similar to the qRFD, however, utilizing only the eigenvalues of the reduced density matrices.

For instance, if $\eta^{\ell}_{\uparrow(\downarrow)}$ and  $\tau^{\ell}_{\uparrow(\downarrow)}$ are the eigenvalues of reduced density matrices $\rho^{\ell}_{0}$ and $\rho^{\ell}(t)$ respectively, and are arranged in the ascending (descending) order then,
\eq{ M_{\ell}(\eta^{\ell}_{\uparrow}, \tau^{\ell}_{\downarrow}) \leq F_{\ell}(\rho^{\ell}_{0}, \rho^{\ell}(t)) \leq M_{\ell}(\eta^{\ell}_{\uparrow}, \tau^{\ell}_{\uparrow})\label{fidelitybounds},}
where, $M_{\ell}(\eta^{\ell}_{\uparrow}, \tau^{\ell}_{\uparrow}) = \left(\sum_{p=1}^{2^{\ell}}\sqrt{(\eta^{\ell}_{\uparrow})_{p} (\tau^{\ell}_{\uparrow})_{p}}\right)^{2}$ is the fidelity between the initial and quenched diagonal states. 
From the upper bound of the local fidelity, 
we propose the minimum reduced fidelity distance,
\eq{d_{\ell}^{m}(t) = -\log(M_{\ell}(\eta^{\ell}_{\uparrow}, \tau^{\ell}_{\uparrow})). \label{mRFD}} 
%

This distance is an effective measure as seen in subsequent sections to study and distinguish the quenches that cause minimal changes in long-range correlations in the system after the quench shown in Fig. \ref{MI_Int}(a) - (d) and have avoided crossings in the reduced density spectrum.
Unlike the quantum reduced fidelity distance, the minimum reduced fidelity distance is a finite-size measure without any thermodynamic counterpart; therefore, as the $\ell \rightarrow L \rightarrow \infty$, such distances will necessarily become zero. This, however,  is particularly suited for experiments as local observables are relatively easily accessible in quantum simulators. 
\section{Results}
In this section, we show that irrespective of the nature of the quench, the qRFD serves as a local measure for dQPT, as observed in both integrable and non-integrable models.
We also explain the category of quenches in which mRFD approximates the critical time and exponent.
\subsection{dQPT in integrable Ising model}
We consider the one-dimensional (1D) transverse field Ising model~(TFIM) with the following Hamiltonian, 
\eq{H = - \sum_{i=1}^{N} \left(\sigma_{i}^{z}\sigma_{i+1}^{z} + h \sigma_{i}^{x}\right) \label{TFIM_spin_Hamiltonian},}
where $\sigma$'s represent the Pauli matrices, $h$, the transverse field strength and $N$, the size of the spin chain.
%
\begin{figure}[tbh]
    \centering
    \includegraphics[width=1\columnwidth]{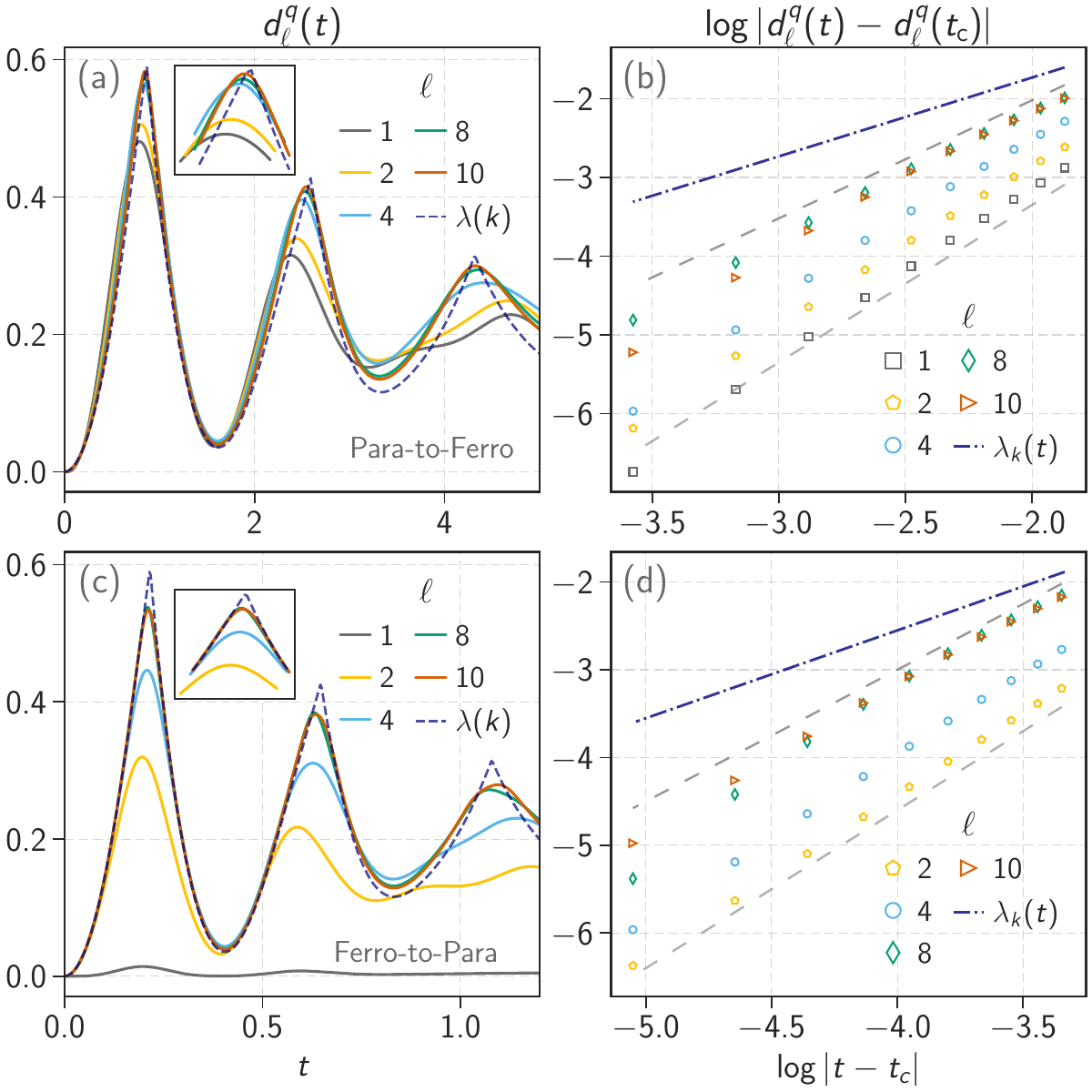}    \caption{{\bf qRFD for TFIM.} (a) and (b) corresponds to quench-I with $h_{i}=4$ and $h_{f}=0.25$. (c) and (d) relates to quench-II with $h_{i}=0.25$ and $h_{f}=4$.~In (a) and (c), the qRFD, along with rate function in momentum space \eqref{returnprobabilitymomentumspace} is shown. The inset shows a magnified area near the first critical point, where the peaks of qRFD approach that of the rate function as the subsystem size $\ell$ increases. (b) and (d) show the scaling analysis of qRFD in both quenches. The blue dotted line is the scaling of $\lambda_{k}(t)$ which have a slope equal to one. We observe the exponent value to reduce significantly from $1.95 (1.85)$ to $1.53(1.49)$ as subsystem size increases from $\ell=2$ to $L/2$ for quench-I (quench-II). The system size for this calculation is $L = 20$.}
    \label{Int_qrfd}
\end{figure}
\begin{figure}[!tbh]
    \centering
    \includegraphics[width=1\columnwidth]{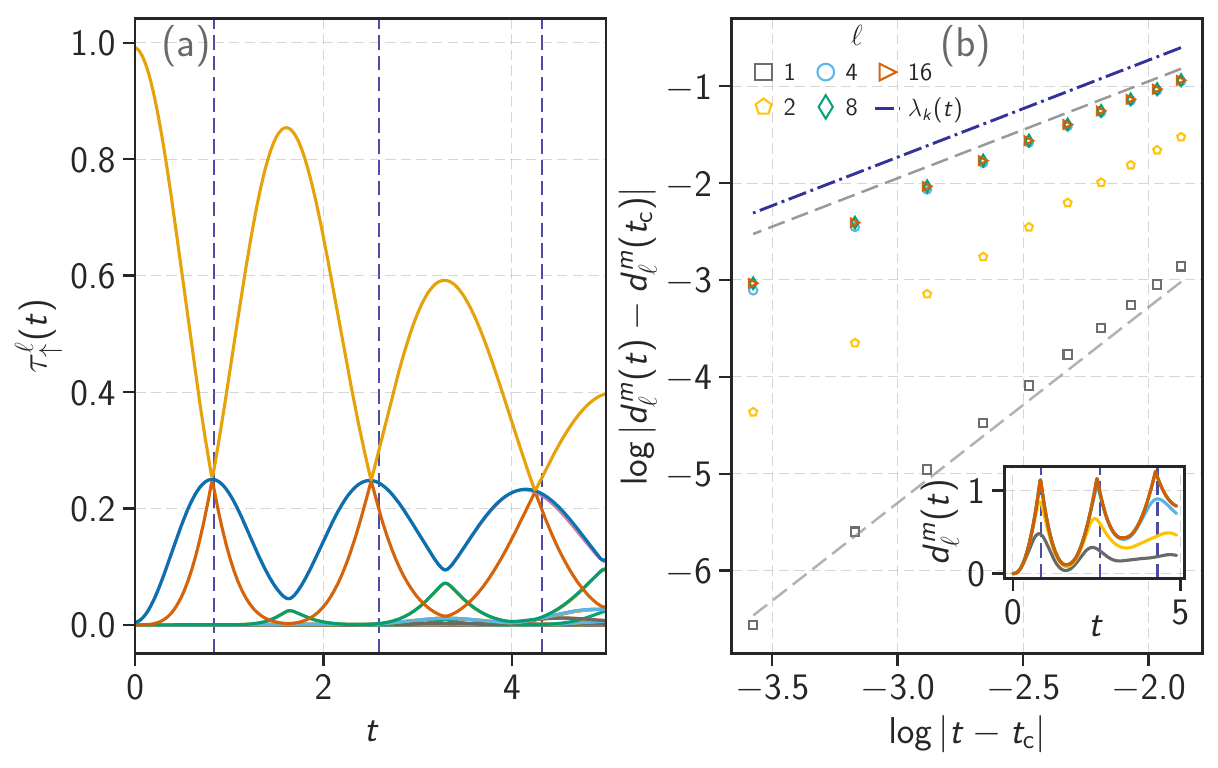}
    \caption{{\bf mRFD for TFIM.} (a) shows the reduced density spectrum of $\rho^{\ell}(t)$, corresponding to quench-I with $h_{i}=4,~h_{f} = 0.25$ and $\ell=8,~L=64$. The blue dotted lines represent the critical times. Close to the critical times, the spectrum exhibits avoided crossing. The inset in (b) is the mRFD for the same quench parameters. The dotted lines correspond to the critical time calculated using Eq.~\eqref{criticaltime_TFIM}. (b) shows the scaling analysis of mRFD near the first critical point. The slope varies between $2.02$ for $\ell=1$ to $1.02$ for $\ell=8,16$. Blue dotted line represents the scaling of rate function.
    } 
    \label{Int_rdsp_mrfd}
\end{figure}
The corresponding momentum space Hamiltonian reads, 
\eq{H(k) = \Vec{d}(k,h).\Vec{\sigma} = 2\sin{k} \sigma^{y} + 2 (h- \cos{k})\sigma^{z},\label{TFIM_momentum_hamiltonian}}
with $\Vec{d}(k,h) = (0,~2\sin{k},~2(h- \cos{k}))$ and $\Vec{\sigma} = (\sigma^{x},~\sigma^{y},~\sigma^{z})$.
The equilibrium model shows a quantum phase transition from paramagnetic to ferromagnetic (vice-versa) when $h$ changes across the critical point $h_{c} = 1 (-1)$.
The quenches across $h_{c}$ lead to dynamical quantum phase transitions~\cite{Heyl_DQPT_review,Heyl_DQPT,DQPT_Non_Int}.
The rate function can be obtained in momentum space owing to the translation symmetry of the model, 
\eq{\lambda_{k}(t) = -\frac{1}{\pi} \text{Re}(\log \int_{0}^{\pi} dk (|g_{k}|^{2} + \exp^{-2 i \epsilon_{k}^{f}t } |e_{k}|^{2}),\label{returnprobabilitymomentumspace}}
when the system is quenched from $H_{\text{initial}}(k) = \Vec{d_{i}}(k,h_{i}).\Vec{\sigma}$ to $H_{\text{final}}(k) = \Vec{d_{f}}(k,h_{j}).\Vec{\sigma}$ with
$|g_{k}|^{2} = \frac{1}{2}( 1+ \hat{d_{i}}(k).\hat{d_{f}}(k))$,~$|e_{k}|^{2} = \frac{1}{2}( 1- \hat{d_{i}}(k).\hat{d_{f}}(k))$ and $\epsilon_{k}^{f} = |\Vec{d_{f}}(k)|$ \cite{klambda}. 
The rate function exhibits non-analyticity at the critical time,
\eq{t_{c} = \frac{\pi}{\epsilon_{k^{*}}^{f}}\left(n + \frac{1}{2}\right),~n=0,1,2,3....,\label{criticaltime_TFIM}}
where $k^{*} = \cos^{-1}\left[{\frac{h_{i}h_{f}+1}{(h_{i} + h_{f})}}\right]$~\cite{klambda,Heyl_DQPT_review,Heyl_dqpt_TFIM,Heyl_survey}.
The scaling analysis yields the following power-law scaling close to the transition, 
\eq{\lambda_{k}(t) - \lambda_{k}(t_{c}) \sim \left(\frac{t-t_{c}}{t_{c}}\right)^{\nu},\label{Heylscaling1d}}
with an exponent $\nu=1$ in 1D \cite{Heyl_scaling_dqpt}.
In 2D the scaling law is logarithmic, $\lambda(t) - \lambda(t_{c}) \sim (t-t_{c})^{2}\log|t-t_{c}|$.
See App.~\ref{2d TFIM} for more details on the scaling of the $\lambda(t)$ in 2D Ising model.  
%

\subsubsection{qRFD for quench-I and quench-II}\label{secintegrablemodel}
The qRFD~\eqref{qRFD} identifies the critical time and the exponent in dQPT observed under both quenches.
The system is initialised in a paramagnetic phase ($|h_{i}|>1$)~and quenched to ferromagnetic phase ($0<|h_{f}|<1$) in quench-I, while the quench-II is from the ferromagnetic to paramagnetic phase. 

Figure \ref{Int_qrfd}(a) and (c) show the quantum reduced fidelity distance \eqref{qRFD} for different subsystem sizes $\ell$ for quench-I and quench-II, respectively.
The initial reduced density matrix $\rho_{0}^{\ell}$ and time evolved reduced density matrix $\rho^{\ell}(t)$ are evaluated exactly.
The rate function \eqref{returnprobabilitymomentumspace} corresponding to these parameters is shown in Fig.~\ref{Int_qrfd}(a) and (c).
The qRFD faithfully approaches the rate function $\lambda(t)$ at the critical time when the subsystem size increases to $\ell \rightarrow L/2$, where $L$ is the system size. 
\footnote{Though the subsystem size is taken continuously in all the data shown in the main text, the qRFD averaged over a few combinations of random partitions also identifies the transition times as shown in Appendix~\ref{Random_section}. }
In this particular quench setup, the post-quench correlations are local; therefore, even a subsystem as small as a single spin can also signal the approximate transition time as seen in Fig.~\ref{Int_qrfd}(a).
However, from Fig.~\ref{Int_qrfd}(c), a minimum of $2$ spins are required to characterize the critical time for quench-II, where the quench results in development of long-range correlations.

In Fig.~\ref{Int_qrfd}(b) and (d), the scaling analysis of qRFD and the rate function near the critical point for two quenches is shown. 
Similar to $\lambda_{k}(t)$, the qRFD obeys the expected power law scaling,
$$
d_{\ell}^{q}(t) - d_{\ell}^{q}(t_{c}) \sim (t - t_{c})^{\nu}, \label{Int_qrfd_scaling}
$$
where $\nu \sim 1.95 (1.85)$ for $\ell = 2 $ and $\nu \sim 1.53 (1.49)$ for $\ell = 10$ for paramagnetic to ferromagnetic (ferromagnetic to paramagnetic) quench.  
We observe a steady decrease in the exponent value towards the expected Ising universality value $\nu \approx 1$ as the subsystem size $\ell$ increases towards $L/2$.
The observed discrepancy is related to limited access to large system sizes, which we address through mRFD in the next section. 
\subsubsection{mRFD for quench-I}\label{mrfd_and_Correlations} 
The mRFD \eqref{mRFD} distinguishes and characterizes the dQPT resulting from quench-I, which has nearest neighbour dominated spatial entanglement structure after the quench as seen in Fig.~\ref{MI_Int}.
For the integrable model, the initial $(\eta_{\uparrow}^{\ell})$ and quenched $(\tau_{\uparrow}^{\ell}(t))$ reduced density spectrum and hence the mRFD is calculated from single particle correlation matrix~\cite{Peschel}.
For details of the calculation, refer to App.~\ref{correlationcalculation}.
Unlike qRFD, this allows us to go to a larger system sizes with the following condition $\ell/L << 1$.
We start discussing the evolution of eigenvalues $\tau_{\uparrow}^{\ell}(t)$ of reduced density matrix $\rho^{\ell}(t)$ for quench-I as shown in Fig.~\ref{Int_rdsp_mrfd}(a). 
The spectrum exhibits avoided crossings near the critical time of dQPT.
The inset in Fig.~\ref{Int_rdsp_mrfd}(b) shows the mRFD \eqref{mRFD} calculated from the reduced fidelity between initial and time evolved diagonal states with eigenvalues, $\eta^{\ell}_{\uparrow}$ and $\tau^{\ell}_{\uparrow}(t)$.
The blue dotted lines in the inset represent the critical time calculated using Eq.~\eqref{criticaltime_TFIM}.
With increasing subsystem sizes $\ell$, the mRFD approaches the critical time.
Nonetheless, we observe deviation from the exact critical time in the avoided crossings (critical time) in the reduced density spectrum (mRFD) as seen in Fig.~\ref{Int_rdsp_mrfd}(a) and (b). 
However, it vanishes for the quench-I with $h_{i}=4$ and $h_{f}=0$,  where the return amplitude peaks do not decay and the avoided crossings in the reduced density and mRFD shows non-analyticity precisely at the critical time calculated using Eq.~\eqref{criticaltime_TFIM}. 
Such discrepancies are also observed in Ref.~\cite{DQPT_entanglement3} for topological models under the quenches with oscillatory degeneracies in entanglement spectrum, and is believed to be due to specific nature of the temporal correlations in the subsystem.

The scaling analysis of mRFD near the critical point gives,
$$
d_{\ell}^{m}(t) - d_{\ell}^{m}(t_{c}) \sim (t - t_{c})^{\nu}, \label{scaling_mrfd_int}
$$
with the critical  exponent $\nu \sim 1$ for $\ell = 8\, \text{and}\, 16$, which is what one would expected and as seen in the rate function Fig.~\ref{Int_rdsp_mrfd}(b).
Hence, though mRFD is a finite size observable, for quench-I in TFIM, it serves as an efficient local quantity to understand the scaling and universality of dQPT.
For quench-II, where the change in long-range correlations is significant, the information in diagonal states is insufficient to understand the dQPT fully.
Like entanglement entropy, the mRFD exhibits linear growth near the critical time for quench-II. 

Here we considered the dQPT observed during quenches across the equilibrium critical point in 1D TFIM.
The efficaciousness of qRFD and mRFD in more general scenarios, such as understanding anomalous dQPTs, where the quenches are done within the same equilibrium phase, and 2D Ising model is discussed in Appendix \ref{Anomalous} and \ref{2d TFIM} respectively.
 %
\begin{figure}
    \centering
    \includegraphics[width=1\columnwidth]{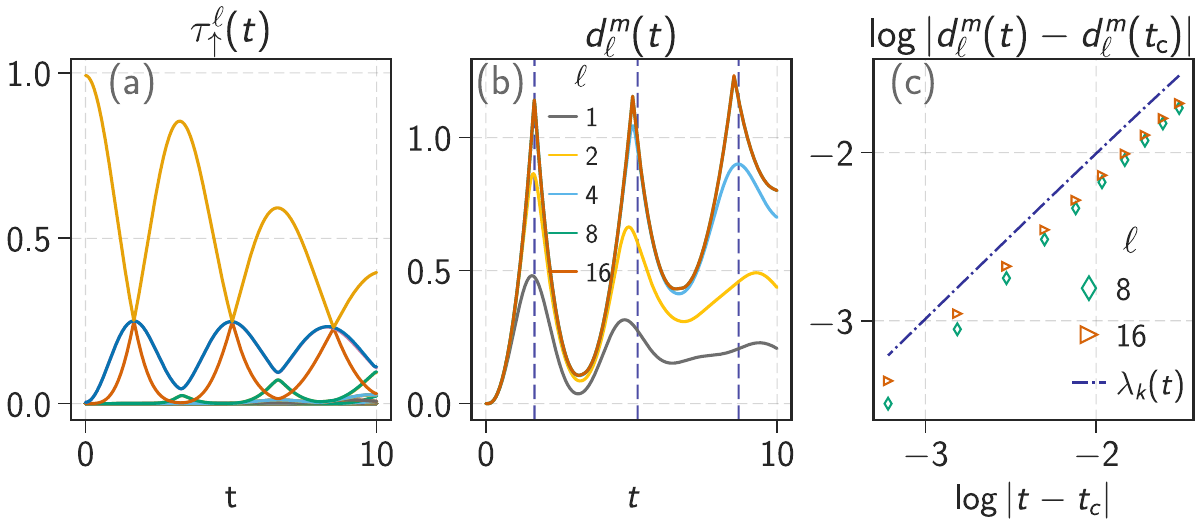}
    \caption{{\bf mRFD for topological model.} (a) and (b) is the reduced density spectrum and mRFD calculated for the quench-I. The spectrum is for $\ell=8$ and $L=64$. The scaling analysis of mRFD gives a critical exponent $\nu \sim 0.96$ for $\ell=8$ and similarly for $\ell=16$. The blue dotted line in (c) represents the scaling of return amplitude calculated in momentum space.}
    \label{A3_mrfd}
\end{figure}
%
\subsubsection{mRFD and topological dQPT} \label{topologyandmrfd}
The fermionic momentum space Hamiltonian in Eq.~\eqref{TFIM_momentum_hamiltonian} has a well-defined topology.
The topological order parameter, winding number,~$\nu_{D} = \frac{1}{2 \pi i}\int_{-\pi}^{\pi} dk \frac{d}{dk} \log(h(k)),~\text{where}~h(k) = d_{x}(k)-i d_{y}(k)$ identifies the trivial and topological phases~\cite{Asboth}.
The quench-I in the spin chain \eqref{TFIM_spin_Hamiltonian} corresponds to a quench from the trivial $(\nu_{D}=0)$ to topological phase $(\nu_{D}=1)$ in the momentum space Hamiltonian. 
Under this quench, the correlations are local while the presence of edge states in the initial phase contributes to long-range correlations for the quench-II~\cite{SSH_DQPT}. 
Hence, the mRFD is a useful measure for detecting the occurrence of a topological dQPT in fermionic models under quench-I.

Figure~\ref{A3_mrfd}(a), (b), and (c) show the quenched reduced density spectrum, mRFD, and its scaling for the topological chiral symmetric fermionic model under quench-I.
Appendix \ref{A3} includes a detailed description of the model and the efficiency of qRFD in capturing the critical time and exponent of dQPT resulting from quench-I and quench-II.
In the vicinity of the critical point of quench-I, avoided crossings and sharp kinks are seen in the quenched reduced density spectrum and the mRFD, respectively.
The mRFD displays much faster convergence to the expected power law scaling  with an exponent $\nu \sim 0.95$ for $\ell = 8$ compared to qRFD.
\subsection{dQPT in non-integrable model}\label{secnonintegrable}
We consider the following non-integrable transverse field next-nearest neighbour Ising spin chain,
\eq{ H = -\sum_{i=1}^{N} h \sigma_{i}^{x} + \sigma_{i}^{z}\sigma_{i+1}^{z} + \Delta \sigma_{i}^{z}\sigma_{i+2}^{z}.\label{Non_Int_SpinHamiltonian}}
Depending on the strength of integrability breaking interaction, $\Delta$ and the transverse field strength $h$, the equilibrium model exhibits four distinct phases: paramagnetic, ferromagnetic, floating phase and an anti-phase \cite{DQPT_Non_Int, ANNI}.
The dQPT arises when any or both parameters, $\Delta$ and $h$, are quenched across an equilibrium critical point~\cite{DQPT_Non_Int, quasilocal}.
It is observed that the increase in the absolute value of $\Delta$, the integrability breaking term, along the positive (negative) direction manifests as a decrease (increase) in the critical time scale.
Unlike the critical time, $t_{c}$ \eqref{criticaltime_TFIM} of the TFIM, there is non-equal spacing of $t_{c}$ when $\Delta \neq 0$ \cite{DQPT_Non_Int}.
\begin{figure}[!tbh]
    \centering
    \includegraphics[width=1\columnwidth]{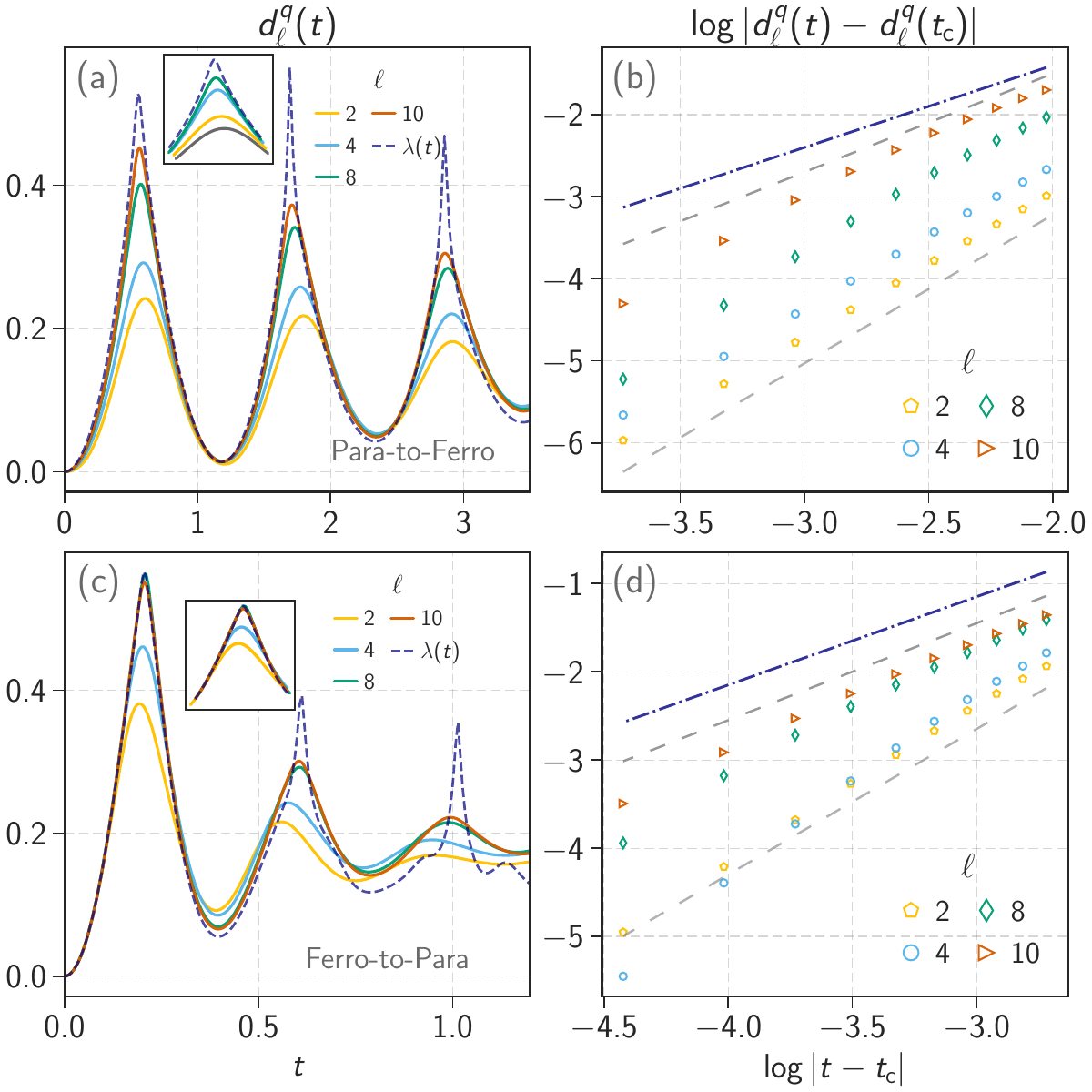}
    \caption{{\bf qRFD for non-integrable model.} The qRFD \eqref{qRFD} for different subsystem sizes and the real space rate function \eqref{returnprobability} is shown in (a) and (c). The quench parameters are $h_{i}=2(0),~\Delta_{i}=0.5(0)$, $h_{f}=0.5(4)$,~$\Delta_{f}=0.5(0.6)$ for quench-I (quench-II). (b) and (d) show the scaling analysis of qRFD near the first critical point. The slope of the curve varies from $1.75(1.65)$ for $\ell = 2$ to $1.20(1.10)$ for $\ell=10$ for quench-I (quench-II). The blue dotted lines are reference curves with the slope one. The system size is $L=20$.}
    \label{qrfd_Non_Int}
\end{figure}
%
\subsubsection{qRFD for quench-I and quench-II}\label{secnoninte_quenches}
We follow the similar philosophy of the previous section to choose parameters such that it represents the quench-I and quench-II protocols and shows dQPT.
Instead of momentum space, here, we calculate all the density matrices directly in the real space. This severely limits the accessible system sizes in exact diagonalization calculations.
Figure~\ref{qrfd_Non_Int}(a) and (c) show the quantum reduced fidelity distance \eqref{qRFD} along with the rate function for quench-I and quench-II.
As the subsystem size, $\ell$, increases, the qRFD approaches the critical point at which the rate function shows non-analyticities in time.
The scaling analysis of qRFD near the first critical point, shown in Fig.~\ref{qrfd_Non_Int}(b) and (d), indicates a similar scaling law; the qRFD follows a power law scaling, $d_{\ell}^{q}(t) - d_{\ell}^{q}(t_{c}) \sim (t - t_{c})^{\nu}$, with $\nu \sim 1.75 (1.65)$ for $l = 2$ and $\nu \sim 1.20 (1.10)$ for $\ell=10$ for quench-I (quench-II). 
The blue dotted line in Fig.~\ref{qrfd_Non_Int}(b) and (d) is the curve that has a power law behavior with exponent one.

For this particular model, \textcite {quasilocal} showed that for a chain with $L = 16$ spins and a local string size ranging from $10 \leq \ell \leq L$, the quasilocal observable exhibits power law scaling with an exponent $\nu \sim 1$.
%
Similarly, as the subsystem size $\ell$ increases, the critical exponent of qRFD of the non-integrable model gradually approaches that of the qRFD of the integrable model, but with  visible finite-size effects.
%

\begin{figure}
    \centering
    \includegraphics[width=1\columnwidth]{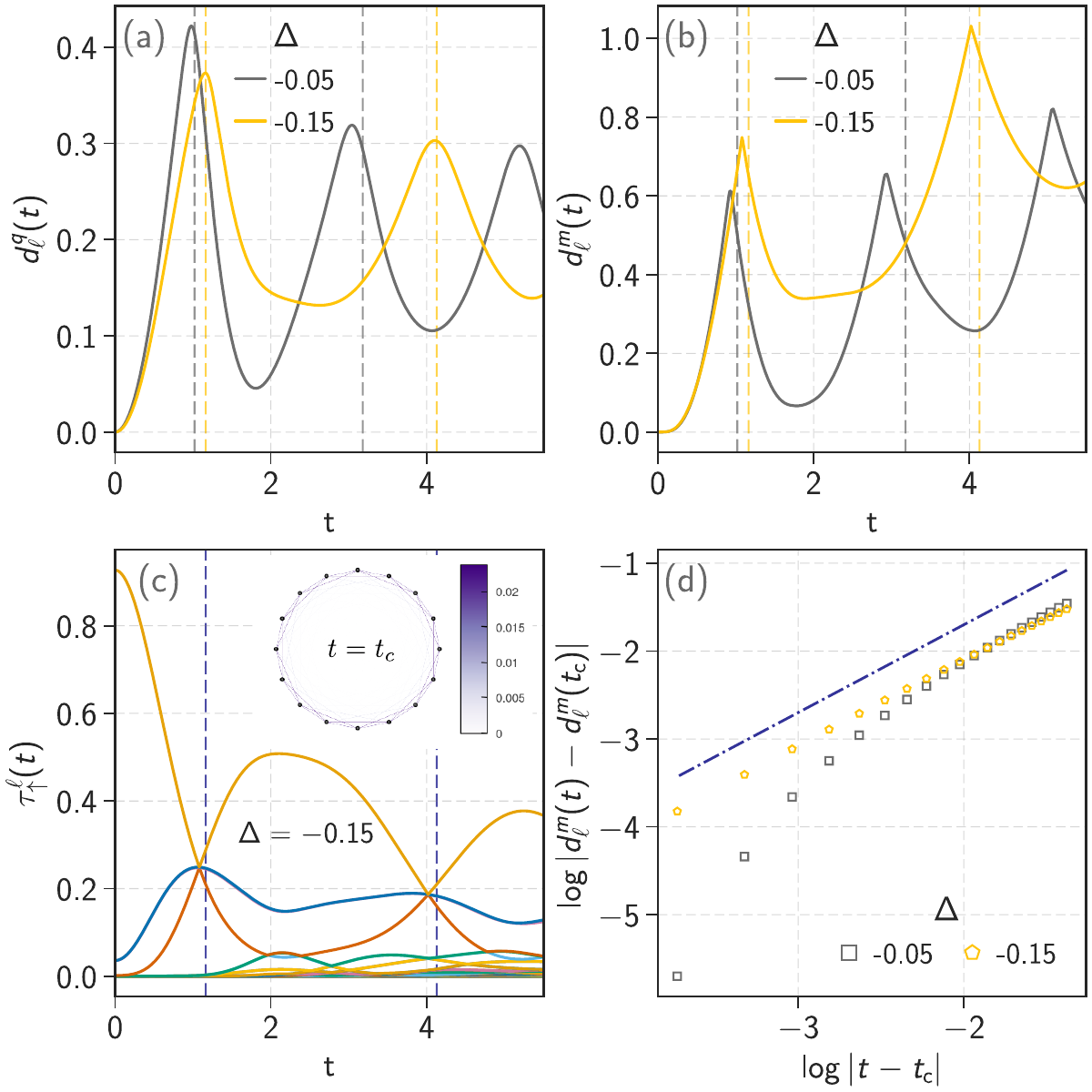}
    \caption{{\bf mRFD for non-integrable model} (a) and (b) show the qRFD~\eqref{qRFD} and the mRFD~\eqref{mRFD} for quench-I for different $\Delta$,~$h_{i} = 1.3$,~$h_{f}=0.2$,~$\ell=10$, and $L=20$. The non-analyticities of mRFD displays a shift compared to the transition time of qRFD, and it depends on $\Delta$. The reduced density spectrum for quenched state corresponding to $\Delta=-0.15$ is shown in (c).  The inset represents the change in mutual information after the quench. (d) Represents the scaling analysis for the mRFD for $\Delta={-0.05, -0.15}$. For $\Delta=-0.05$ the slope is $1.17$ and for $\Delta=-0.15$ it is $0.95$ for subsystem size $\ell=10$. The blue line indicates a slope of one.}\label{Non_int_mrfd}
    \end{figure}

\subsubsection{mRFD for quench-I}
Figure~\ref{Non_int_mrfd}(a) and (b) are the qRFD and mRFD corresponding to quench-I with varying $\Delta$ ($\Delta_{i}=\Delta_{f}=\Delta$), with dotted lines representing critical time obtained from the rate function.
The critical time scale of qRFD and mRFD increases upon increasing the strength of integrability breaking term, $\Delta$. 
No simple quantitative relation exists for $t_{c}$ like the 1D TFIM, and the dQPT appears at unequal time intervals.
Similar to the integrable case, the transition time of the quench-I dQPT in mRFD experiences a shift dependent on $\Delta$ relative to the critical time of the rate function or qRFD.
Figure ~\ref{Non_int_mrfd}(c) shows the reduced density spectrum with avoided crossing near the critical time. 
The inset in Fig.~\ref{Non_int_mrfd}(c) reveals that, even in the presence of integrability breaking terms, the mRFD identifies and distinguishes quench-I, scenarios where post-quench changes in quantum correlations are local.
Figure \ref{Non_int_mrfd}(d) shows the scaling analysis of mRFD for different $\Delta$'s.
The mRFD obeys the power law scaling,~$d_{\ell}^{m}(t) - d_{\ell}^{m}(t_{c}) \sim (t-t_{c})^{\nu}$,~with $\nu \approx 1.17$ for $\Delta=-0.05$ to $\nu \approx 0.95$ for $\Delta=-0.15$.
\section{Conclusion}\label{conclusion}
In this study, we have demonstrated the efficacy of the quantum reduced fidelity distance as a local observable for understanding the dQPT that occurs in diverse models, including integrable and non-integrable spin chains and topological systems. 
The quantum distance metric is defined based on the reduced fidelity between the initial and quenched states for finite subsystems. 
Notably, as the subsystem size $(\ell \sim L/2)$ increases, the critical time and the critical exponents identified by the qRFD gradually converge to the point where the rate function exhibits nonanalytic behavior.
Moreover, the local observable derived from the diagonal states, minimum reduced fidelity distance, distinguishes dQPTs arising from different quenches.
The rationale for introducing mRFD comes from the observation that for a specific quench protocol (quench-I), the entanglement spectrum features avoided level crossings near the transition time during the time evolution. 
Additionally, the quantum correlations developed after quench-I remain local. 
Another key insight is that irrespective of the nature of the quench, the fidelity between diagonal states or eigenvalue distributions sets an upper limit on the quantum reduced fidelity.
These findings underscore the importance of local observables, particularly mRFD, in effectively probing and characterizing the diverse dynamical phases that arise during quantum quench processes.

While the qRFD and mRFD have proven valuable in comprehending dQPT by exhibiting cusps at critical times, some areas will benefit from further improvements.
Firstly, due to limited access to large system sizes, the critical exponent, $\nu$, obtained from the scaling analysis of qRFD does not precisely match with the expected exponent $\nu\simeq 1$.
However, the relative error in calculating the exponent, $\delta \nu = \nu -1$, decreases from $0.95$ to $0.53$ ($0.75$ to $0.20$)  upon increasing the subsystem size from $\ell=2$ to $\ell = L/2$ for the integrable (non-integrable) spin model.
The finite size effects observed for non-integrable model are less compared to the integrable model. 
Interesting to note that similar finite size effects are also observed in the real-space local effective free energy proposed by~\textcite{Local_measures}.
Though the critical time is reliably found from the effective free energy corresponding to subsystem size as small as $\ell=2$, the non-analytic behavior becomes more evident as the subsystem size increases to $\ell \geq 32 $. 
In this respect, mRFD effectively addresses the finite size effects in integrable models by calculating the eigenvalues of the reduced density matrix from the single-particle correlation matrix, enabling access to large system sizes. 
For instance, we successfully identified the critical time and precise critical exponent for a subsystem size of $\ell=8$ for $L=64$. 

Finally, note that with the emergence of techniques capable of measuring reduced density matrices, the qRFD and mRFD can serve as convenient methods for experimentally detecting dQPTs~\cite{Expt_reduced_density, Expt_reduced_fidelity2, Expt_reduced_density2}.
For instance, the local projective measurements were performed on a subset of Bose-Einstein condensate of \ce{^{87}Rb} atoms in an optical lattice \cite{expt_reduced_density_finite_time} and the investigation of the propagation of non-local correlations in the trapped \ce{^{171}{Yb}^{+}} ions \cite{expt_reduced_density_finite_time2} provides information about the subsystem density matrix at finite time and thus exemplifies the practical applicability of the qRFD and mRFD.

\section*{acknowledgement}
We would like to  thank Himadri S. Dhar for an earlier collaboration on a closely related project, and several useful discussion. We thank Markus Heyl for a critical reading of the manuscript. We also gratefully acknowledge the valuable discussions with the late Amit Dutta during this reserach.  SB would like to thank MPG for funding through the Max Planck Partner Group at IITB. RB would like to thank DST-INSPIRE fellowship No. IF180067 for funding. We use open-source \texttt{QuSpin} package~\cite{Quspin1,Quspin2} for exact diagonalization calculation.

\bibliography{ref}

\setcounter{equation}{0}
\setcounter{figure}{0}
\setcounter{table}{0}
\renewcommand{\thefigure}{S\arabic{figure}}

\appendix
\section{Reduced density spectrum from two-point correlation function for integrable model}\label{correlationcalculation}
The initial ground state of Eq.~\eqref{TFIM_momentum_hamiltonian}  at $h_{i} = 4$,
\eq{\ket{\psi_{k}} = u_{k} c_{k}^{\dagger} c_{-k}^{\dagger} \ket{0} + v_{k} \ket{0},\label{groundstateising}}
is used to calculate the one-body correlation matrix,
\eq{\mathbf{C}_{\text{initial}} = \begin{pmatrix}
\mathbb{C} && \mathbb{F}\\
\mathbb{F}^{\dagger} && \mathbb{I} - \mathbb{C}
\end{pmatrix},\nonumber}
where,
\eq{\mathbb{C}_{ij} = \langle c_{i}^{\dagger} c_{j} \rangle = \frac{2}{L}\sum_{k \in BZ/2} |u_{k}|^{2} \cos(k(i-j)),\nonumber \\ 
\mathbb{F}_{ij} = \langle c_{i}^{\dagger} c_{j}^{\dagger} \rangle = \frac{2}{L} \sum_{k \in BZ/2}  u_{k}^{*} v_{k} \sin(k(i-j)),\nonumber}
and $i,j$ runs over the subsystem size $\ell$.
The entanglement spectrum ($ \epsilon$) is calculated from the eigenvalues of correlation matrix ($\zeta$),
\eq{ \zeta_{q}^{\ell} = \frac{1}{e^{\epsilon_{q}^{\ell}} + 1},\nonumber}
where $q \in 0,1,...\ell$.
The $2^{\ell}$ eigenvalues of initial reduced density matrix ($\rho_{0}^{\ell}$) is then obtained from the entanglement spectrum as \cite{boundaryfidelityssh},
\eq{\eta_{p}^{\ell} = \frac{1}{\mathcal{Z}} \Pi_{q} e^{-\epsilon_{q}^{\ell} n_{q}^{(p)}},\nonumber}
where $n_{q}^{(p)} \in {0,1}$ are single level occupation numbers. 
$\mathcal{Z}$ is the normalisation constant such that $\sum_{p} \eta_{p}^{\ell} = 1$.
To observe dQPT, the ground state \eqref{groundstateising} is time evolved with respect to the Hamiltonian \eqref{TFIM_momentum_hamiltonian} at $h_{f} = 0.25$.
At any instant of time, the quenched state is, $\ket{\psi_{k}(t)} = u_{k}(t) c_{k}^{\dagger} c_{-k}^{\dagger} \ket{0} + v_{k}(t) \ket{0}$.
From these time-dependent coefficients, the correlation matrix ($\mathbb{C}(t))$ and hence the quenched reduced density spectrum with eigenvalues arranged in ascending order, $\tau^{\ell}_{\uparrow}(t)$ is calculated using the same procedure described above.
%
\begin{figure}[!tb]
    \centering
\includegraphics[width=1\columnwidth]{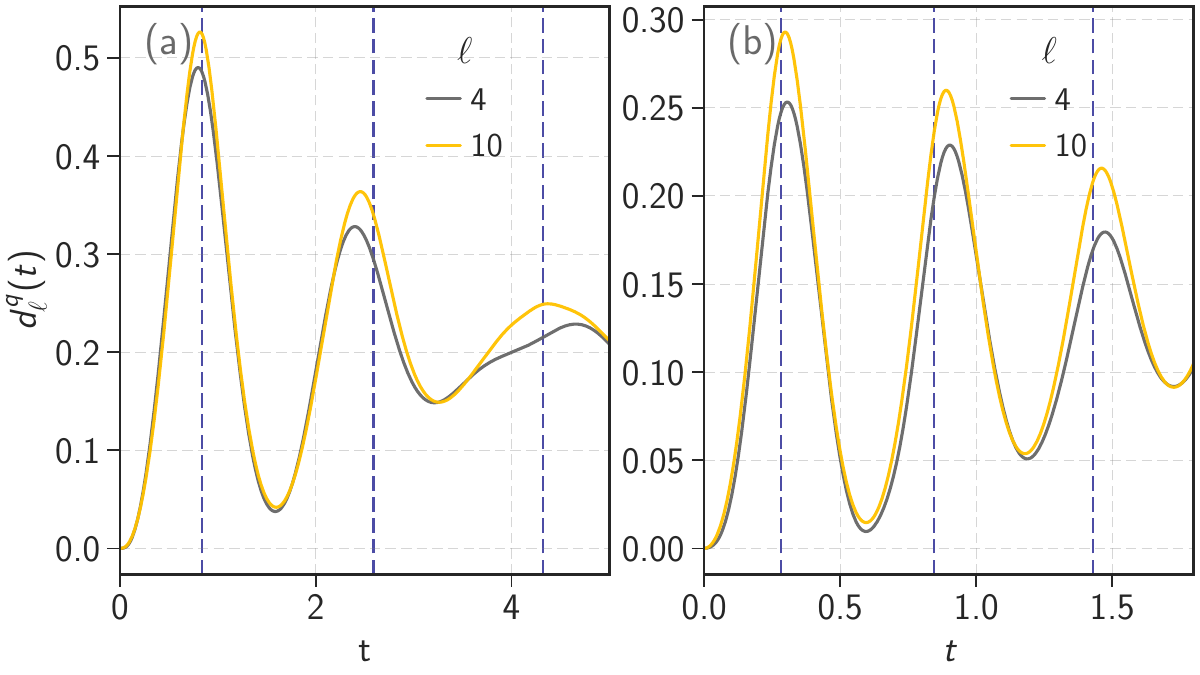}
\caption{{\bf qRFD for random partitions.}(a) and (b) is the plots of the qRFD averaged over ten random partitions of size $\ell=4,10$ for integrable and non-integrable spin chains. The quench parameters chosen for integrable \eqref{TFIM_spin_Hamiltonian} and non-integrable \eqref{Non_Int_SpinHamiltonian} model are $h_{i}=4.0$, $h_{f}=0.25$ and $h_{i}=2,~h-{f}=0.5,~\Delta_{i}=\Delta_{f}=0.5$ respecively. The blue dotted lines represent the critical time. As $\ell$ increases, the qRFD approaches the critical time.} 
    \label{Random}
\end{figure}
%
\begin{figure}
    \centering
\includegraphics[width=1\columnwidth]{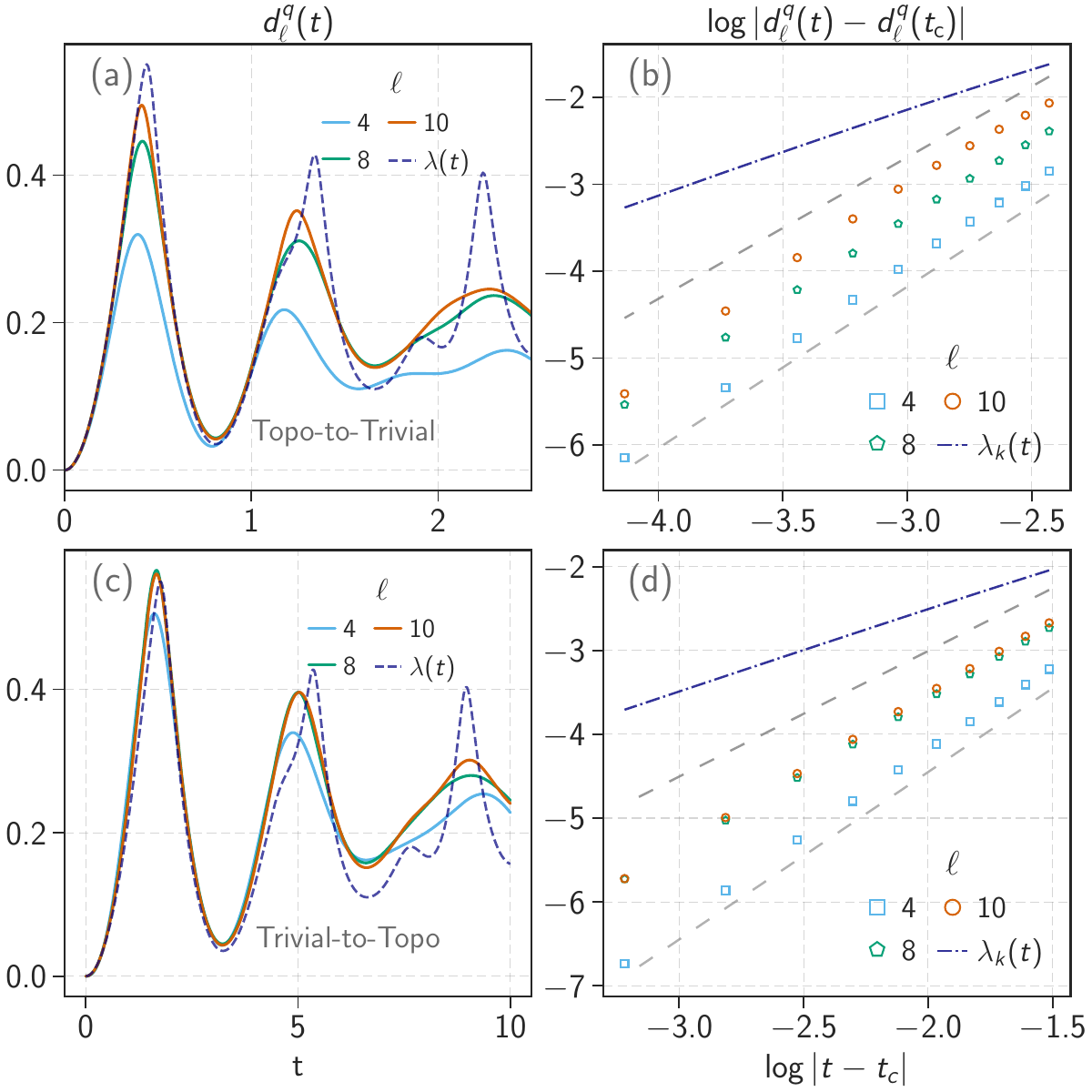}
    \caption{{\bf qRFD for topological model.}(a) and (c) is the qRFD calculated for different subsystem sites for quench-I and quench-II. It also has the rate function calculated in real space. The total number of sites is $L=20$. The quench parameters are $m_{i}=0.25,~m_{f}=4.0$ for Fig.~(a) and $m_{i}=4,~m_{f}=0.25$ for (c).~The scaling analysis of qRFD for both quenches is studied in (b) and (d). The slope of the curve changes from $1.86(1.98)$ for $\ell=4$ to $1.63(1.74)$ for $\ell = 10$ during quench-II (quench-I).~The blue dashed line in (b) and (d) is the slope one curve.}
    \label{A3_qrfd}
\end{figure}
\section{qRFD for random partitions}\label{Random_section}
Here we show the qRFD calculated for dQPT observed in both integrable \eqref{TFIM_spin_Hamiltonian} and non-integrable \eqref{Non_Int_SpinHamiltonian} spin chains when the choice of subsystem sites is random.
Consider the paramagnetic to ferromagnetic quench discussed in Sec.~\ref{secintegrablemodel} and Sec.~\ref{secnoninte_quenches}.
We consider ten configurations of random subsystem sites of size $\ell=4,10$ for $L = 20$.
The qRFD \eqref{qRFD} averaged over these random configurations for the integrable and non-integrable models are plotted in Fig.~\ref{Random}(a) and (b), respectively.
The blue dotted lines represent the critical time.
As the subsystem size increases, the qRFD averaged over random partitions approaches the critical time.
For the opposite quench also, qRFD averaged over random partitions picks up the critical time efficiently.
\section{Fermionic topological model}\label{A3}
\begin{figure}
    \centering    \includegraphics[width=1\columnwidth]{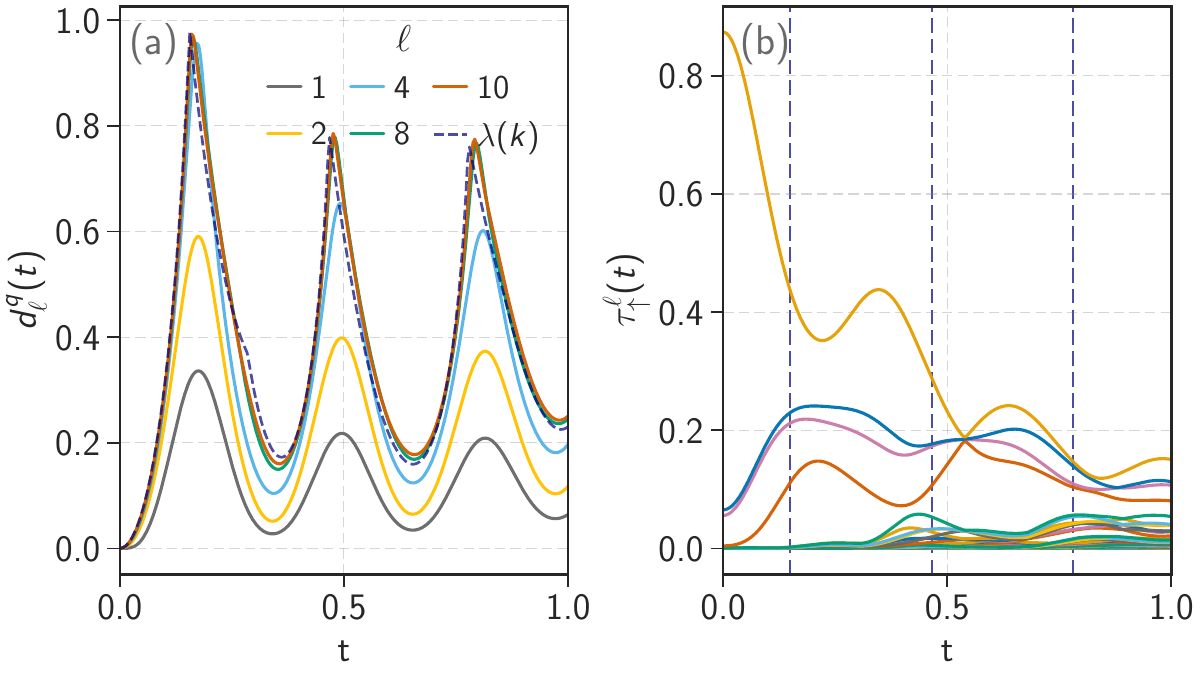}
    \caption{{\bf qRFD for anomalous dQPT.}(a) shows the qRFD plotted for different subsystem sizes and the rate function. The quench performed is from $\gamma_{i}=1.6,~h_{i}=1.5$ to $\gamma_{f}=-4,~h_{f}=3$. The total system size is $L=20$. (b) represents the reduced density spectrum calculated from the single particle correlation matrix for the same quench. The subsystem size is $\ell=8$, and the total system size is $L=64$. The blue dotted lines indicate the critical time at which the rate function diverges.  }
    \label{Anomalous_qrfd}
\end{figure}
%
We consider the chiral symmetric Hamiltonian in Class AIII of topological insulators and superconductors~\cite{A3},
\eq{ H = \sum_{i=1}^{N} (\frac{1}{2} c_{i}^{\dagger}(\sigma^{x} + i \sigma^{y})c_{i+1} + h.c)+ \sum_{i=1}^{N} m~c_{i}^{\dagger} \sigma^{y} c_{i},\label{A3fermionHamiltnonian}}
where $c_{i}^{\dagger} (c_{i})$ are the fermion creation (annihilation) operators, $\sigma$'s are the Pauli matrices, $m$ is the complex dimerisation amplitude and $N$ is the number of unit cell.
In the Fourier space,
\eq{H(k) = \vec{d(k)}.\vec{\sigma} = \cos{k}\sigma^{x} + (m-\sin{k})\sigma^{y}, \label{A3momentumHamiltonian}}
where $\vec{d(k)} = (\cos{k},(m-\sin{k}),0)$ and $\vec{\sigma} = (\sigma^{x},\sigma^{y},\sigma^{z})$.
In equilibrium, the model has a non-trivial topological index for $0<m<1$ and a trivial topological index for $m>1$.
Quenching the parameter $m$ across the equilibrium critical point $m_{c}=1$ results in dynamical quantum phase transitions.
By changing $m_{i}=0.25 (4)$ to $m_{f}=4 (0.25)$, we get topological (trivial) to trivial (topological) quench.

Figure \ref{A3_qrfd}(a) and (b) correspond to the qRFD \eqref{qRFD} calculated for both quenches for different subsystem sizes.
The rate function calculated in real space is also plotted for both quenches.
The qRFD captures the critical point efficiently.
The scaling analysis of both quenches shows a power law behavior,~with the critical exponent $\nu \sim 1.86 (1.98)$ for $l=4$ to $\nu=1.63(1.74)$ for $l=10$  for topological to trivial (trivial to topological) quench.
The mRFD and its scaling for the quench from trivial to the topological phase where only local correlations are present \cite{SSH_DQPT}, is explained in Sec.~\ref{topologyandmrfd}.
 \section {qRFD in anomalous dQPT}\label{Anomalous}
 When the quench parameters do not cross the equilibrium quantum critical point, we get the anomalous dQPT~\cite{DQPT_pure4}.
 The spin Hamiltonian, 
 \eq{H = \sum_{i=1}^{N} -h \sigma_{i}^{z} + J_{x}\sigma_{i}^{x}\sigma_{i+1}^{x} + J_{y} \sigma_{i}^{y} \sigma_{i+1}^{y},\label{spin_xy_Hamiltonian)}}
 exhibits anomalous dQPT.
 Since this is an integrable model, the diagonalization in momentum space is possible with,
 \eq{H(k) = \Vec{d}(k,h).\Vec{\sigma} = 2\gamma \sin{k} \sigma^{y} + 2 (h- \cos{k})\sigma^{z},\label{XY_momentum_hamiltonian}}
where $\Vec{d}(k,h) = (0,~2\gamma\sin{k},~2(h- \cos{k}))$,~$\Vec{\sigma} = (\sigma^{x},~\sigma^{y},~\sigma^{z})$ are the Pauli matrices, $J_{x} = (1 +\gamma)/2$ and $J_{y}= (1-\gamma)/2$.
Consider the quench, $\gamma_{i}=1.6,~h_{i}=1.5$ to $\gamma_{f}=-4,~h_{f}=3$, which is within the paramagnetic phase~\cite{DQPT_pure4}.
Figure \ref{Anomalous_qrfd}(a) corresponds to the qRFD \eqref{qRFD} for different subsystem sizes.
The return amplitude is calculated in the momentum space using Eq.~\eqref{returnprobabilitymomentumspace}. 
As the subsystem size increases, the maximum value of qRFD moves towards the peak value of the rate function, thus identifying the critical time efficiently.

The reduced density spectrum calculated using the single-particle correlation matrix, as explained in Appendix \ref{correlationcalculation}, is shown in Fig.~\ref{Anomalous_qrfd}(b).
Though the spectrum shows a single avoided crossing at some time, it lies much away from the critical time indicated by the blue dotted line and does not show oscillations like Fig.~\ref{Int_rdsp_mrfd}(a).
Hence the mRFD fails to capture the anomalous dQPT.
%
\begin{figure}
    \centering
    \includegraphics[width=1\columnwidth]{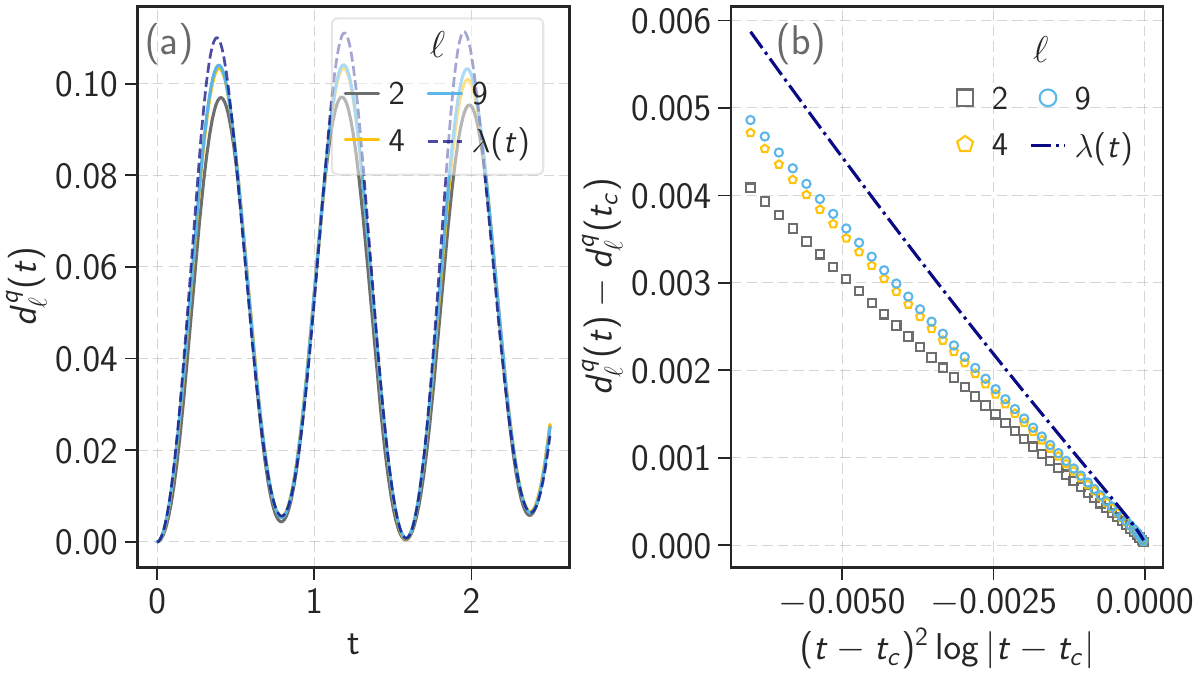}
    \caption{{\bf qRFD for 2D TFIM.}(a) shows the qRFD for different subsystem sizes and $\lambda(t)$ calculated in real space for a $4~\times~4$ square lattice. The system is quenched from $h_{i}=1.5J$ to $h_{f}=0.25J$ for $J=J_{p}=1$. The subsystem size chosen is $\ell=2$ sites and areas with size $\ell = 4 (2 \times 2)$ and $\ell = 9(3 \times 3)$.~(b) represents the scaling analysis of qRFD for different subsystem sizes. As the subsystem size increases,  the qRFD also shows similar logarithmic scaling as $\lambda(t)$.}
    \label{qrfd_2d}
\end{figure}
\section{qRFD in TFIM on a square lattice} \label{2d TFIM}
Consider the TFIM \eqref{TFIM_spin_Hamiltonian} on a square lattice,
 \eq{H = -\sum_{\langle ij \rangle} J_{ij} \sigma_{i}^{z}\sigma_{j}^{z} - h\sum_{i=1}^{L} \sigma_{i}^{x},\label{2dsquarelattice}}
where the nearest neighbour hopping takes value $J_{ij} = J$ along the rows and $J_{ij} = J_{p}$ along the columns, $L$ is the total number of lattice sites.
During the quench from $h_{i} = 1.5J$ to $h_{f}=0.25J$ and for $J=J_{p}=1$, the model undergoes dynamical quantum phase transition.
Near the critical point, the rate function obeys the following scaling relation \cite{Heyl_scaling_dqpt},
\eq{\lambda(t) - \lambda(t_{c}) \sim (t-t_{c})^{2}\log|t-t_{c}|,\label{scaling2d}}
where $t_{c}$ is the critical time.
The quantum fidelity distance for different choices of subsystem size, along with the rate function, is shown in Fig.~\ref{qrfd_2d}(a).
The critical point is identified well by the local quantum distance.
The Fig.~\ref{qrfd_2d}(b) shows that similar to the scaling of $\lambda(t)$ \eqref{scaling2d}, the qRFD also exhibits logarithmic scaling,
\eq{ d_{\ell}^{q}(t) - d_{\ell}^{q}(t_{c}) \sim (t-t_{c})^{2}\log|t-t_{c}| \label{qrfd_2d_scaling}.} 

The reduced density spectrum does not show avoided crossings at the critical time for this quench. Hence, the minimum reduced fidelity distance \eqref{mRFD} is not a reliable local measure.
The quench at which mRFD can capture dQPT in this model needs to be studied further. 
%

\end{document}